\documentclass[aps,pra,twocolumn,preprintnumbers,amsmath,amssymb,floatfix,groupedaddress]{revtex4}

\usepackage[dvips]{graphicx}
\usepackage{dcolumn}
\usepackage{bm}
\usepackage{amssymb}
\usepackage{xr}

\usepackage{color}

\usepackage[dvips]{graphicx}

\begin{document}

\title{Observation and measurement of  ``giant" dispersive optical non-linearities \\ in an ensemble of cold Rydberg atoms.  }

\author{ Valentina Parigi, Erwan Bimbard,  Jovica Stanojevic}

\affiliation{Laboratoire Charles Fabry,~Institut d'Optique,~CNRS,~Univ Paris-Sud,  2 Avenue Fresnel, 91127 Palaiseau, France. }

\author{ Andrew J. Hilliard }

\altaffiliation{Present address : QUANTOP, Institut for Fysik og Astronomi, Aarhus Universitet, Ny 
Munkegade 120, 8000 Aarhus C, Denmark}
\affiliation{Laboratoire Charles Fabry,~Institut d'Optique,~CNRS,~Univ Paris-Sud,  2 Avenue Fresnel, 91127 Palaiseau, France. }

\author{\\ Florence Nogrette, Rosa Tualle-Brouri, Alexei Ourjoumtsev }
\author{ Philippe Grangier }

\affiliation{Laboratoire Charles Fabry,~Institut d'Optique,~CNRS,~Univ Paris-Sud,  2 Avenue Fresnel, 91127 Palaiseau, France. }

\date{\today}


\begin{abstract}
We observe and measure dispersive optical non-linearities in an ensemble of cold Rydberg atoms placed inside an optical cavity. The experimental results are in agreement with a simple model where the  optical non-linearities are due to the progressive appearance of a Rydberg blockaded volume  within the medium. The measurements allow a direct estimation of the ``blockaded fraction" of atoms within the atomic ensemble. 
\end{abstract}


\maketitle

The realization of non-linear optical effects that are large enough to effect photon-photon interactions would be a significant step forward for quantum information processing and communications. In particular, a giant dispersive and non-dissipative non-linearity could enable the implementation of a two-photon phase gate. 
 It is well known that standard optical non-linearities, even the largest ones that are typically resonant $\chi^{(3)}$ effects, are  
too small to reach this range.  Presently, two main approaches have been considered to  reach the  desired regime of deterministic photon-photon interactions. One is cavity (or circuit) QED, where it is experimentally well-established that the atom-field coupling can be large enough to produce single atom - single photon interactions \cite{cavityQED}. However, in order to use such effects for optical ``flying qubits", the challenge is to get very high input-output coupling to the  cavity \cite{in_out}.  Another approach, inspired by recent work on both dark-state polaritons  \cite{polaritons}  and on Rydberg blockade \cite{blockade,rydbergs} is to temporarily convert the photons into other particles, such as Rydberg polaritons, that may have very strong interactions \cite{adamsprl105,kuzmich,vuletic}.

\begin{figure}
\includegraphics[scale = 0.93]{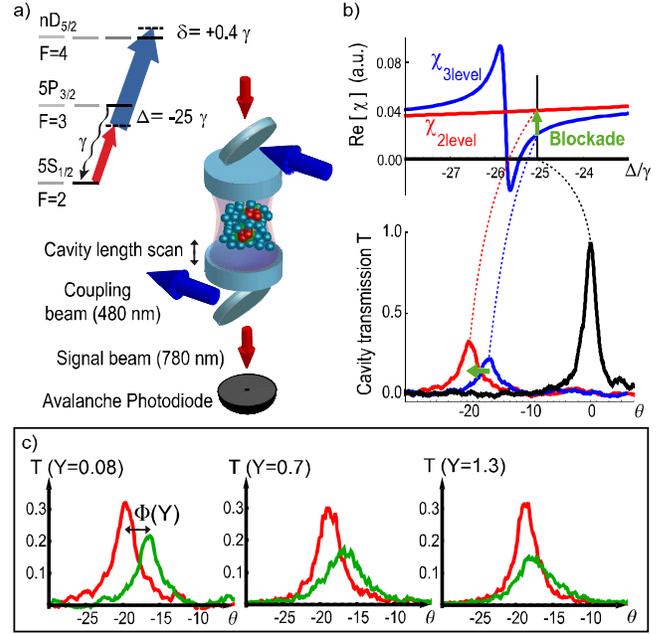}
\caption{a) An ensemble of $N$ three-level atoms inside an optical cavity is excited by a strong (blue) coupling  field, and a weak (red) signal field, all detuned from resonance with respect to the $5S_{1/2} \rightarrow 5P_{3/2} \rightarrow nD_{5/2}$ transitions in Rubidium 87. b) Principle of the measurements. Lower part: black: cavity scan without atoms, red: without the blue beam, blue: with blue light at very low red intensity. $\theta$ is the signal laser-cavity detuning (in units of cavity linewidth), scanned with the cavity length. Upper part: corresponding real part of susceptibilities. The effect of blockade is indicated by the green arrows. c) Measured transmission versus $\theta$ with the coupling field on (green) or off (red) for different normalized red intensities $Y$ in the case of the $n=61$ state. The differential peak shift, denoted by $\Phi(Y)$, is reduced by the blockade for increasing intensities (see text).}
\label{fig:scheme}
\end{figure}

Here, we pursue this approach by using an ensemble of cold Rydberg atoms to create ``giant" dispersive non-linearities on a weak ``signal" beam.
Specifically, we use  atoms in a 3-level ladder configuration (see Fig.~\ref{fig:scheme}a)
driven by a strong (blue) laser beam, detuned from resonance on the upper transition, and a very weak (red) signal
beam on the lower transition. Our scheme is similar to those used in previous work on non-linearities in three-level systems \cite{ghost,qnd}; however, even at optimal performance, the optical non-linearity produced in those schemes was not large enough to be useful at the single-photon level. Here, the two-photon transition involves Rydberg states in order to
exploit their very large van der Waals interactions to further enhance the non-linearity. 

It should be noted that a signifiant challenge in such experiments is that, while the non-linearities are expected to be extremely large at very small intensities, they also quickly saturate with increasing signal beam intensity, whereupon one enters the dipole blockade regime. As such, one must be able to work with very small light intensities and to accurately detect small non-linear phase shifts.  

In this letter, we present the first measurements and a simple physical interpretation of this  ``giant"  dispersive  non-linear effect. At the lowest (non-linear) order  in signal intensity, a  $\chi^{(3)}$ behaviour  is expected, creating intensity-dependent phase shifts.
In order to both create and detect such phase shifts, we use the following experimental approach.  First, in order to increase the dispersive effects with respect to the absorptive ones, the signal and  control fields are detuned from all levels (intermediate and Rydberg, see Fig.~\ref{fig:scheme}a). Second,  the atoms are located inside a low-finesse optical  cavity, in order  to amplify the effects while keeping a high input-output coupling efficiency. Since the cavity is itself an interferometer, it converts the non-linear phase shift into a shift of the cavity resonance peak; the position of this peak can be measured as a function of the intensity of the signal  light, with and without the coupling blue light. 

Theoretically, the system is described by the Hamiltonian for a 3-level atomic ensemble
in the presence of blue and red laser fields, and the van der Waals interaction potential between Rydberg atoms. This leads to a hierarchy of Bloch equations containing atom-atom correlation operators \cite{jovica,pohl2}. The susceptibility $\chi$ experienced by the signal field is determined by the atomic coherence of the lower transition of the ladder system. As we are interested in its value at low field intensities, the hierarchy may be truncated to second order (two-body correlations), leading to a closed set of equations  \cite{pohl2}. The solution of these equations may be performed numerically for the full three-level model, or analytically by appropriate adiabatic eliminations. Both cases recover the result obtained in \cite{pohl2} for the particular case of zero two-photon detuning. 

Additionally, this analysis must take into account that the blue and red light have different beam waists, standing wave structures and very different finesses (respectively 2 and 120) in the optical cavity. Hence, the spatial variations of the two fields must be introduced in the numerical evaluation of Bloch equations, and averages performed over the intensity distributions. 
In addition, the coupling of the injected cavity mode  with other transverse cavity modes, induced by the non-linear term of susceptibility, might  produce losses and additional line-shifts \cite{courty}, but this effect was calculated to be negligible in our experimental conditions.

The numerical and analytical solutions we obtained  for the dispersive part in the $\chi^{(3)}$-limit confirm a modified version of the ``universal scaling" introduced in 
 \cite{pohl}, where the susceptibility of the medium is expressed as:
\begin{equation}
\chi = \chi_{3level}  + p_b \;  (\chi_{2level} - \chi_{3level}). 
\label{us}
\end{equation}
Here,  $\chi_{2level}$  is the susceptibility of the lower one-photon transition without blue light,  $\chi_{3level}  $  is the susceptibility of the same transition with blue light but without Rydberg-Rydberg interactions, and $p_b$  is the probability for an atom to be blockaded 
due to the Rydberg-Rydberg interaction. 
The intuitive explanation of the non-linear effect is the following: if we inject a very weak red signal beam in the presence of the blue light on the two-photon transition, it will experience the single-atom 3-level dispersive phase shift ($p_b \sim 0, \chi \sim \chi_{3level}) $. 
As the red intensity is increased, the Rydberg state population will increase with the effect that each excited Rydberg atom will detune
from the two-photon resonance all neighbouring atoms inside a blockade sphere, because of the Rydberg-Rydberg interaction. 
Therefore the 3-level component of the dispersion will be reduced, and the dispersion of the medium will go back towards its value in absence of blue light ($p_b \rightarrow 1 $, and thus $\chi \rightarrow \chi_{2level}  $).
In equation (\ref{us}),  $\chi_{2level}  $ and  $\chi_{3level} $ 
can be obtained from standard 2-level and 3-level optical Bloch equations (without the Rydberg interaction term), while $p_b$ must be
inferred from the full model.
The result is that, to lowest order in the red beam intensity, one can write the simple relation (holding for a homogeneous system)
\begin{equation}
p_b = (p_3 - p_{3_{coll} } )/ p_3 = n_b \; p_3,
\label{pb}
\end{equation}
where $p_3$  is the Rydberg population without Rydberg-Rydberg interactions, $p_{3_{coll}}$ is the Rydberg population in presence of interactions, and  $n_b $ is the number of atoms in a blockade sphere \cite{pohl2}, more precisely defined by
\begin{equation} n_b =  (2 \pi^2/3) \rho \sqrt{|C_6| / \delta_{e}}.
\label{nb}
\end{equation}
In this expression,  $\rho$ is the atomic density, $C_6$ is the usual coefficient in the van der Waals  interaction \cite{blockade,rydbergs}, and  $\delta_{e}$ is  the two-photon detuning, corrected  (and actually increased, see Appendix) by the blue-induced light shift.

The experimental scheme is shown in Fig.~\ref{fig:scheme}a.  
A cloud of cold $^{87}$Rb atoms in a magneto-optical trap is placed into an optical  cavity, with finesse  $F \sim 120$ and linewidth $\kappa/2\pi\sim$ 10 MHz at 780 nm. The atomic sample is cooled to 40 $\mu$K by 6 ms of polarization gradient cooling, whereupon the sample is optically pumped to the $5S_{1/2} (F=2,m_F=2)$ state. Approximately 1 nW of 780 nm light is coupled into the cavity; the red light is  detuned by $\Delta=-75$ MHz $\sim - 25 \gamma$  below the  $ 5 S_{1/2} (F=2,m_F=2) \rightarrow  5 P_{3/2} (F=3,m_F=3) $ transition with linewidth $\gamma$.  The cavity length is scanned around resonance with the red beam. Approximately 100 mW of a 480 nm beam is also injected into the cavity, using  a dichroic mirror. The two beams are slightly blue-detuned from the two-photon transition toward the Rydberg state 
$ n D_{5/2} (F=4,m_F=4)$, with $n=46,\,50,\,56,\,61$. On the cavity output side, the blue and red beams are separated by a second dichroic mirror,  and the 780 nm light is focused on an avalanche photodiode (APD). Since the two-photon detuning has to be rather small (typically 1 MHz), the locking system of the lasers should be designed to ensure a narrow linewidth of the two-photon transition. For this purpose, the red (780 nm) and blue (480 nm) lasers are locked onto the same ``transfer" cavity, as well as a far detuned laser (810 nm)  locking  the experimental cavity.

The choice of the sign of the detunings is very important, because neither  the light shifts nor the Rydberg interactions should  bring the atoms (or pairs of atoms) into resonance, otherwise the losses become very high. This is both predicted theoretically \cite{pohl2} and observed experimentally.
We are using $nD_{5/2}$  Rydberg states,  that have attractive interactions and may involve many different potential curves.  One reason for this is to fulfill the above condition on  the sign of the detunings. The increased risk to create ions in the cloud due to attractive Rydberg-Rydberg interactions is discussed below. The value of $\sqrt{C_6}$ is calculated by averaging  $\sqrt{C_6(\lambda)}$ over potentials $U_{\lambda}$ of a given $nD_{5/2}+nD_{5/2}$ manifold, where $\lambda$ enumerates the molecular states. These effective $C_6$ lie within 20 $\%$ from the values calculated in \cite{raithel}.

The cavity resonance position corresponds to a measured value of $\theta = (\omega-\omega_c)/\kappa$, $\omega$ and $\omega_c$ being the red laser and cavity frequencies ($\theta=0$ is the resonance position without atoms).
In the absence of blue light and well below one-photon saturation, the atoms induce a shift of this position proportional to $C \chi_{2level} \sim C \gamma 
/\Delta$. It depends on  the detuning $\Delta$ and on the cooperativity parameter $C=Ng^{2}/2 \gamma\kappa$ 
where $g$ is the usual atom-field coupling parameter, and $C$ takes into account  the collective enhancement due to the $N$ atoms within the cavity mode. In the presence of blue light, the shift becomes $\propto C \chi$ where $\chi$ is the general susceptibility, expected to be theoretically defined by (\ref{us}) in our parameter range.
We measured the blue-induced part of the resonance shift $\Phi(Y)$ (see Fig.~\ref{fig:scheme}c), where $Y$ is the red intensity normalized to the saturation intensity at resonance on the lower transition. A significant $Y$-dependent blue-induced resonance shift at low values of $Y$ is an indication of the desired collisional non-linear dispersive effect. It is convenient to normalize this shift to its value for vanishing red power, and to consider $\Phi(Y)/\Phi(0)$.
From (\ref{us}) and (\ref{pb}) its theoretical value is given by:
\begin{equation}
\frac{\Phi(Y)}{\Phi(0)}=\frac{(\chi-\chi_{2level})}{(\chi_{3level}-\chi_{2level})}=1-n_{b}p_{3}.
\label{deltateta}
\end{equation}
To lowest order, $p_3 \propto Y$, so the quantity $1-n_{b}p_{3}$ should manifest a $Y$-dependence and, according to (\ref{nb}), a Rydberg-level dependent behaviour. It should be noted that $p_3$ is also dependent on the blue Rabi frequency (see Appendix) and, since the different Rydberg levels possess different dipole moments, a given blue power corresponds to different values of Rabi frequency for the different $n$ states. However, after averaging over spatial intensity distributions as discussed above, the averaged $\overline{p_{3}}$ is only very weakly sensitive to the state-dependent variation of blue Rabi frequency in our parameter range.
\begin{figure}
\includegraphics[scale = 1.1]{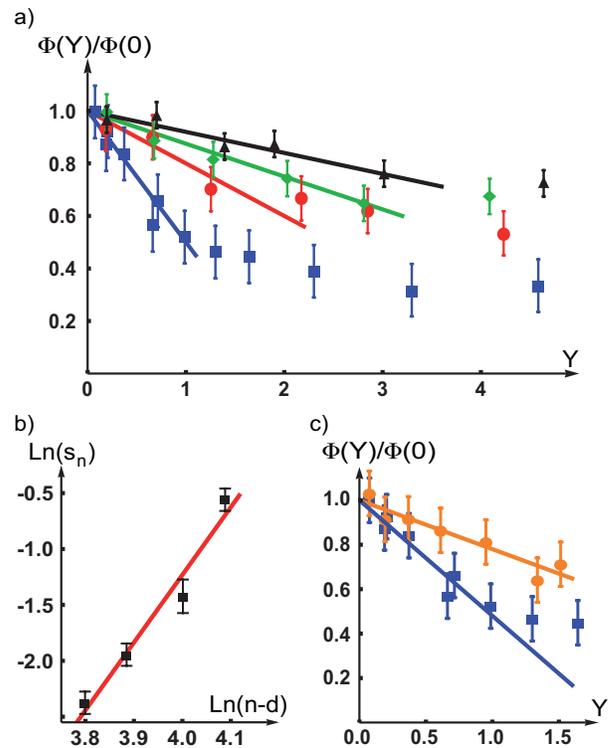}
\caption{a) Normalized cavity shift $\Phi(Y)/\Phi(0)$ versus normalized input red intensity $Y$, for several Rydberg states with $n=$ 46 (black triangles), 50 (green diamonds), 56 (red circles), 61 (blue squares). The full lines corespond to the initial slopes of the curves, which are expected to be of the form $1 - s_n Y$. b) Value of $s_n$ as a function of $(n-d)$, in logarithmic scales. The slope of this curve gives the expected power law behaviour. c) Normalized cavity shift $\Phi(Y)/\Phi(0)$ as a function of the normalized input red intensity $Y$, for the Rydberg state $n=61$, for two different atomic densities $\rho_{low} \sim$ 0.02 at/$\mu$m$^3$ (orange circles), and $\rho_{high} \sim $0.04 at/$\mu$m$^3$ (blue squares). The observed change in slope is consistent with the expected ratio $\rho_{high}/\rho_{low} \sim 2$ of the densities.}
\label{fig:results}
\end{figure}

Figure~\ref{fig:results}a shows the measured shift  versus $Y$ for Rydberg states with  different $n$. Below a certain $Y$ value, the behaviour is well described by the function $1-s_n Y$. As the red power increases, a saturation effect appears: the number of atoms in the Rydberg level stops  increasing proportionally to the red intensity. If our simple description is correct, the initial slope $s_n$  should be proportional to  $\sqrt{C_6}$ which scales as  $\sqrt{n^{*11}}=n^{*5.5}=(n-d)^{5.5}$ where $d \approx 1.35$ is the quantum defect \cite{blockade,rydbergs}. However, $s_n$ actually contains an additional contribution from the intrinsic non-linearity of $\chi_{3level}$. The measured $s_n$, corrected for this calculated (small) contribution, are plotted as a function of $n^{*}$ in log-log scale in Fig.~\ref{fig:results}b. A linear fit yields a slope of $6\pm 0.5$, consistent with the expected 5.5.
To confirm that the observed effect is due to atomic interactions, we show in  Fig.~\ref{fig:results}c the results obtained for $n = 61$, by decreasing the atomic density, and keeping the same cooperativity. This is achieved by loading more atoms in the MOT and letting the cloud expand a longer time, so the same cooperativity is obtained from a larger cloud with less density. In that case, one expects a reduced non-linear collisional effect, since interactions are density dependent ($n_b$ is proportional to $\rho$). The two sets of data points in the figure correspond to the two values $\rho_{low} \sim $ 0.02 at/$\mu$m$^3$, and $\rho_{high}\sim$ 0.04 at/$\mu$m$^3$: the observed change in slope (by a factor $2.3 \pm 0.3$) is consistent with the density ratio $\rho_{high}/\rho_{low} \sim 2$.

It is also important to take into account the time-scale of the experiment. The measurements - the scans of the cavity length around the red resonance - are done in a transient regime: the time it takes to scan one cavity linewidth is approximately 4 $\mu$s, which should be compared to the time necessary to reach the steady state value of the Rydberg population. The two times  turn out to be of the same order of magnitude, so we have used dynamical rather than steady state solutions, leading to some reduction of the observed non-linear effect. The final calculated values of the slopes $s_n$, taking into account spatial averaging and dynamics, agree with the experiment to within 50 $\%$.

An additional complication arises from the fact that the blockade effect could be mimicked by the unintended creation of ions in the medium. Such ions can be generated either from single atoms, due to the interaction of Rydberg atoms with the ambient blackbody radiation,  or from collisions between Rydberg atoms. In the latter, the collisions are enhanced since we work with Rydberg states with attractive atom-atom potentials.  By increasing the laser powers and scanning the cavity at lower speed, we observed typical cascade ions effects \cite{viteau}, which in our case lead to a 
reduction in the cooperativity through atom loss.  However, ions do not seem to play a role in the regime of parameters where we measured $\Phi(Y)/\Phi(0)$. To confirm this, we varied the duration of exposure to the red light during the scans and did not observe significant changes in the non-linear cavity shift, while the number of ions should change dramatically \cite{robicheaux,viteau}. Furthermore, for the  most relevant parameter range, corresponding to the lowest red intensity, the Rydberg state population is actually very small ($< 5\%$), and the number of ions must be even smaller, typically by several orders of magnitude. 

In conclusion, it is interesting to compare the observed $\chi^{(3)}$ with other references. The resonance shift we observe for the $n=61$ Rydberg state corresponds to an effective value of $Re[\chi^{(3)}] \sim 5.10^{-9} m^2/V^2$. This value, which is the first measurement of a dispersive non-linear susceptibility of such magnitude in Rydberg gases,
is approximately two orders of magnitude below the value reported in \cite{adams}, which was for a stronger absorptive on-resonance process. It is worth noting that in our setup the non-linear phase shift corresponding to this $\chi^{(3)}$ is multiplied by the cavity finesse.

It can thus be expected that the kind of non-linearity observed here can be extended to the single photon regime by reducing the cavity beam waist, increasing the cavity finesse and choosing a Rydberg state with higher quantum number $n$. Correspondingly, one should increase the blue intensity in order to keep a large enough blue-induced phase shift, despite  the  decrease of the dipole matrix element as
$n$ increases. A better description of the losses (due to the atoms or to the cavity) also has to be developed.
Though much progress is still needed to reach the regime of large dispersive
photon-photon interactions in the optical domain, in our system the 
interaction-induced non-linearities exceed by several orders of 
magnitude the "usual" non-linearities resulting from a collection of 
one-atom effects.

\vskip 0.3cm

\noindent {\it Acknowledgements.} This work is supported by the ERC  Grant 246669 ``DELPHI''.  We thank Andr\'e Guilbaud and Fr\'ed\'eric Moron for essential help with the experiment. 

\vskip 0.3cm

\noindent {\bf Appendix.} 
We give here explicit expressions of quantities used in the text, obtained from suitable approximations in standard optical Bloch equations. Let us introduce again the number $n_b$ of atoms in a blockade sphere \cite{pohl2}, defined by
\begin{equation} 
n_b = \frac{2 \pi^2   \rho}{3} \sqrt{\frac{|C_6|}{\delta - g_b^2/\Delta}},
\end{equation}
where $C_6$ is the standard van der Waals coefficient, $\delta$ is the two-photon detuning ($\delta>0$), $\Delta$ the one-photon detuning ($\Delta<0$) , and $g_b$ the blue laser Rabi frequency.  To lowest order for our experimental parameters, the Rydberg population $p_3$ without interactions is : 
\begin{equation} 
p_3 =  \frac{g_a^2 g_b^2}{\Delta^2 (\delta - g_b^2/\Delta)^2} \; \frac{ (\gamma_b \Delta^2 + \gamma g_b^2)}{ (\gamma_c \Delta^2 + \gamma g_b^2)},
\end{equation}
where $g_a$ is the red Rabi frequency ($Y\propto g_a^2$), $\gamma_b$ and $\gamma_c$ are the coherence and population damping times of the Rydberg level.
The population $p_{3_{coll}}$ of the Rydberg state in presence of interactions is: 
$p_{3_{coll}} =  p_3 (1 - p_3 \; n_b)$,
and  the real part of the ``differential" susceptibility  is:
\begin{equation} 
(\chi_{3level} - \chi_{2level})  \propto  \frac{g_b^2}{\Delta^2 (\delta - g_b^2/\Delta)}. 
\end{equation}
As noted above, averaging over the spatial distributions of intensities has been carried out for comparison with the experimental data.


\end{document}